\documentstyle[aps,preprint]{revtex}
\tightenlines
\begin{document}

\input epsf

\def\slashchar#1{\setbox0=\hbox{$#1$}           
   \dimen0=\wd0                                 
   \setbox1=\hbox{/} \dimen1=\wd1               
   \ifdim\dimen0>\dimen1                        
      \rlap{\hbox to \dimen0{\hfil/\hfil}}      
      #1                                        
   \else                                        
      \rlap{\hbox to \dimen1{\hfil$#1$\hfil}}   
      /                                         
   \fi}                                         %

\centerline{\bf Lattice Fields and Extra Dimensions}
\medskip

\centerline{{Michael Creutz}}
\centerline {(creutz@bnl.gov)}
\medskip
\centerline{Physics Department}
\centerline{Brookhaven National Laboratory}
\centerline{Upton, NY 11973}
\bigskip
\begin{abstract}
Lattice gauge theory is now well into its third decade as a major
subfield of theoretical particle physics.  I open these lattice
sessions with a brief review of the motivations for this formulation
of quantum field theory.  I then comment on the recent drive of
lattice theorists to include a fictitious ``fifth'' dimension to treat
issues of chiral symmetry and anomalies.
\end{abstract}


\noindent {Keywords: {lattice gauge theory, domain-wall fermions}
\bigskip

Lattice gauge theory, now a mature subject, continues to attract
considerable attention as a first principles ``solution'' of hadronic
physics.  The basic formulation goes back to Wilson's classic 1974
paper\cite{wilson74}.  The subject remained fairly quiet until an
explosive growth in the early 1980's.  The field is currently
dominated by numerical simulations, although there is considerable
opportunity for analytical developments.  We now have an annual
lattice conference, moving around the world and attracting typically
over 300 participants.

The goals of the lattice community are indeed grandiose.  We are
attempting first-principles calculations in non-perturbative field
theories.  Among the most successful targets are direct calculations
of hadronic spectra, weak matrix elements relevant to extracting the
parameters of the standard model, and the parameters of a new phase of
matter, the quark gluon plasma.  The next few talks will expand on
these calculations.  Here I set the stage with a few comments on the
basic formulation, trying to explain why we go to a lattice at all.  I
then turn to some exciting recent developments driving the community
to simulations in more than four space time dimensions.

So, if space is continuous, why do we go to a lattice?  This is
primarily due to two familiar facts.  First is the importance of
non-perturbative phenomena in strong interaction physics.  Quark
confinement is inherently non-perturbative; an interacting theory of
hadrons is qualitatively different from a perturbation on free quarks
and gluons.  Second, quantum field theory is rampant with ultraviolet
divergences requiring regularization.  The issue is that most cutoff
schemes are based on perturbation theory.  You calculate Feynman
diagrams, and when one is infinite you cut it off.  However, Feynman
diagrams are perturbation theory.  It is the need for a
non-perturbative regularization that drives us to the lattice.

If you get nothing else from this talk, remember that the purpose of
our space-time lattice is nothing but a non-perturbative cutoff.  It
is a mathematical trick.  On a lattice there is a minimum wavelength,
given by the lattice spacing $a$; see Fig.~1.  In Fourier space, this
corresponds to a maximum momentum of $\pi/a$.  The scheme gives a
mathematically well defined system, allowing numerical computations.
This last point has come to dominate the field, but at a deeper level
is secondary to providing a definition of the theory.

I now sketch some of the elegant features of Wilson's \cite{wilson74}
original formulation.  The concept of a gauge theory means different
things to different people.  One way to think of a gauge theory is as
a theory of phases.  As it travels through space time, a charged
particle's wave function acquires a phase
\begin{equation}
U_{i,j} \sim \exp (i\int_{x_i}^{x_j} A^\mu dx_\mu)
\end{equation}
where the line integral is along the path of the particle.  From this
point of view, natural lattice variables are phase factors associated
with the links along which a quark hops.  This approach also makes the
generalization to a non-Abelian gauge theory particularly simple; the
phases are replaced with unitary matrices.  For the strong
interactions, on any link connecting nearest neighbors we have a 3 by
3 unitary matrix $U_{ij} \in SU(3)$.  See Fig.~2.  The size of the
matrix, 3, is determined by the empirical spectroscopic fact that
there are 3 quarks in a proton.

For dynamics, we need a field strength analogous to $F_{\mu\nu}$ in
the continuum.  Since this is a generalized curl, we are naturally led
to consider small loops.  Our basic action is a sum over elementary
squares, called``plaquettes''
\begin{equation}
S=\int d^4x F^{\mu \nu} F_{\mu\nu}\longrightarrow 
-{1\over 3} \sum_p {\rm ReTr}(U_p)
\end{equation}
where the four sides of a given plaquette are multiplied as in Fig.~3.
The variable
\begin{equation}
U_p= U_{1,2}U_{2,3}U_{3,4}U_{4,1}
\end{equation}
represents the flux through the corresponding plaquette.

Given our variables and action, we want to do quantum mechanics.  Here
the basic approach is via path integrals.  We exponentiate the action
and integrate over everything
\begin{equation}
Z=\int (dU) e^{-\beta S}
\end{equation}
Since our variables are in a Lie group, it is natural to define $dU$
as the invariant group measure.  The parameter $\beta$ defines the
bare gauge charge
\begin{equation}
\beta={6 \over g_0^2}.
\end{equation}

Numerical simulation has dominated lattice gauge theory for most of
its history.  The algorithms derive from the mathematical equivalence
of our path integral with a partition function in statistical
mechanics.  In this analogy, the link variables correspond to spins,
interacting with a four-spin coupling at a ``temperature'' $1/\beta$.
A computer simulation sweeps through stored configurations of a finite
system.  With pseudo-random numbers, the program makes random changes
biased by the Boltzmann weight.  This evolution proceeds towards a set
of configurations mimicking ``thermal equilibrium''
\begin{equation}
P(C)\sim  e^{-\beta S}
\end{equation}
What is so enticing about this method is that the computer memory
contains the entire configuration; in principle the theorist can
measure anything desired.  Of course it is not always so simple, Monte
Carlo simulations have statistical fluctuations.  Theorists are faced
with the novel situation of having error bars!

In addition to the statistical errors are several sources of
systematic uncertainties.  These include finite volume and finite
lattice spacing corrections.  Furthermore, present algorithms only
work efficiently for heavy quarks, requiring quark mass
extrapolations.  In addition, many simulations make what is called a
valence approximation for the quark fields.  This neglects feedback of
the quark fields on the gluons, saving perhaps two orders of magnitude
in computer time.

A full inclusion of dynamical quark fields remains not completely
understood.  Some problems arise directly from the anti-commuting
nature of fermion fields.  The path integral is no longer a classical
statistical mechanics problem, but involves operator manipulations in
a Grassmann space.  This complication is usually evaded by integrating
the fermionic fields analytically as a determinant via the famous
formula \cite{matthewssalam}
\begin{eqnarray}
Z&&=\int dA\ d\bar\psi d\psi\ \exp(S_g + \bar \psi (\slashchar{D} +m) \psi)\cr
&&=\int dA\ e^{S_g}\ \vert \slashchar{D} +m \vert
\end{eqnarray}
This determinant, however, is of an extremely large matrix, tedious to
calculate.  While many clever tricks have been developed, existing
schemes remain, in my opinion, ugly and awkward.

The fermion issue becomes considerably worse when a chemical potential
is present.  This is the case for studies of a background baryon
density.  Then the determinant is not positive-definite, wreaking
havoc with Monte Carlo methods.  In this case, except for very small
systems or special toy models, no viable simulation algorithms are
known.  This is the primary unsolved conceptual problem in lattice
gauge theory.

In addition to the algorithmic issues, fermion fields raise
fascinating questions in connection with chiral symmetry.  Here the
difficulties are intertwined tied with the so called ``chiral
anomalies'' of quantum field theory.  Of the extensive recent activity
in this area, my favorite approaches involve extending space-time to
more than four dimensions, making our 4d world an interface in 5d.  I
turn to this subject for the remainder of this talk.

To see how these higher-dimensional schemes work, I sneak up on the
problem by studying an amusing ladder molecule in an applied field
\cite{icepaper}.  Consider two rows of atoms connected by vertical,
horizontal, and diagonal bonds, as sketched in Fig.~4.  On such a
lattice, an electron initially placed on one atom will spread through
the lattice much like water poured in a cell of a metal ice cube tray
(Feynman used this analogy in his Caltech lectures of the mid 60's).
Now apply a magnetic field of strength one half flux unit per
plaquette orthogonal to the plane of this molecule.  This introduces
gauge dependent phases on the bonds; one convention for these factors
is shown in Fig.~5.

Through interference effects, the magnetic field inhibits the
spreading of an electron's wave function.  One consequence is a pair
of special states bound on the ends of the ladder.  One corresponding
wave function is shown in Fig.~6.  A symmetric state is bound on other
end; its wave function is obtained by inverting this figure.

Symmetry considerations drive these special states to zero energy.
The ends of the chain are symmetric under a flip of the system around
an axis parallel to the field; so, the end states must have equal
energy $E_L=E_R$.  On the other hand, the overall sign of the
Hamiltonian can be flipped in two steps.  First make a gauge change by
multiplying all fermion operators on the lower side of the ladder by
$-1$.  This changes the signs of all vertical and diagonal bonds.
Then change the signs of the horizontal bonds with a left right
interchange of the ladder ends.  The overall modification implies
$E_L=-E_R$.  The only way both conditions can be true is if the states
are at exactly zero energy.

This symmetry argument shows that these zero modes are robust under
renormalizations.  No fine tuning of bond strengths is required.  If
the chain is not infinitely long, there can be a small, exponentially
suppressed, mixing of these states.  This will result in energies
$E\sim e^{-\alpha L}$, where the parameter $\alpha$ depends on the
details of the bond strengths and $L$ is the length of the chain.

These zero modes lie at the heart of the domain-wall fermion approach
\cite{dwf}.  We promote the spinor components of a Dirac field on each
space time site into a chain as discussed above.  One can imagine the
chain extending into a fictitious ``fifth'' dimension.  The ``zero
modes'' are then interpreted as the physical quarks.  The basic
picture is sketched in Fig.~7.  The robust nature of these zero modes
means that massless fermions remain so when interactions are turned
on.  Any mass renormalization is proportional to the bare mass, with
no additive contributions.  This is precisely the role played by
chiral symmetry in the ``continuum.''

Going back to the ladder analogy, it is easy to see why these modes
are automatically chiral.  For this we first create a ``device'' by
joining one such ladder onto the side of another, as illustrated in
Fig.~8.  In this figure, the side chain is our fifth dimension, while
the straight chain represents one of the physical space-time
dimensions.  We augment the model, replacing the factors of $i$ on the
horizontal spatial bonds with $i$ times a Pauli spin matrix.  The
direction in which a surface mode moves is then determined by the
direction of its spin.  This device is a helicity separator.  For more
details, see Ref.~\cite{icepaper}.

I now remark on the exact symmetries of this domain-wall formalism.
The zero modes require a surface to exist.  If we were to follow the
old Kaluza-Klein \cite {kaluzaklein} picture and curl the fifth
dimension up into a circle, they would be eliminated.  We must cut the
circle somewhere, as in Fig.~9.  If the size of the extra dimension is
finite, the modes mix slightly.  Indeed, this is crucial because
otherwise there would be no anomalies.

For two flavors, I can obtain the needed zero modes by cutting the
circle twice, as in Fig.~10.  Now there is one exact chiral symmetry
coming from the fact that the fifth dimension involves two
topologically distinct pieces.  Following the notation from the
figure, the number of $u_L+d_R$ particles is absolutely conserved, as
is $u_R+d_L$.  In more usual notation, the axial-vector current
\begin{equation}
j_{\mu 5}^3 = \overline \psi \gamma_\mu\gamma_5 \tau^3 \psi
\end{equation}
is rigorously conserved, even with finite $L_5$.  There does, however,
remain a small flavor breaking.  Since they are different components
of the same fields, $u_L$ and $d_R$ will have an exponentially
suppressed mixing.  This existence of one rigorous chiral symmetry and
a small flavor breaking is reminiscent of Kogut-Susskind
\cite{kogutsusskind} fermions; however, now we have the length of the
fifth dimension to control the size of the flavor breaking.

This scheme gives two flavors from a single five-dimensional field.
This naturally leads to speculations about more zero modes and more
complicated manifolds.  Could this be a route to the flavor/family
structure of the standard model?  Fig.~11 sketches a conceptual scheme
for obtaining three colors of quark and a lepton from a single field.

The question mark in this figure must involve some mechanism for the
baryon decay anomaly of the 't Hooft process.  Also, for gauge
invariance, appropriate quantum numbers should be transferred between
the various singularities giving the physical fermions.  One proposed
scheme is sketched in Fig.~12, representing the rendition of the model
of \cite{crtx} as presented in \cite{lat97}.

In conclusion, I hope I have convinced you that the lattice provides a
powerful non-perturbative regularization, allowing controlled
calculations of hadronic processes.  I am occasionally asked if the
lattice might actually be real.  I don't particularly like this
option, which would lead to an uncomfortable flexability.  Requiring a
continuum limit should limit physical results to renormalizable field
theories.  Of course, ultimately experiment will have to decide.

Despite the maturity of the field, old unresolved fermion issues
remain.  The existing algorithms are rather awkward, and none are
known for dealing with a background baryon density.  Chiral gauge
theories remain controversial, but domain-wall fermions appear to be
making progress.  Some more speculative questions on which the lattice
may eventually shed light are whether mirror species\cite{montvay},
such as massive right handed neutrinos, should exist, and whether true
chiral theories must be spontaneously broken, as observed in the
standard model.

I hope I have at least amused you about the helpful nature of extra
dimensions for chiral symmetry.  There is some hope that similar
techniques can give natural lattice formulations of super-symmetry; an
intriguing scheme \cite{susy} has been proposed for a lattice
formulation of super-symmetric Yang-Mills theory, where the low energy
spectrum has all masses protected from fine tuning.  Of course, the
use of extra dimensions also suggests connections with the recent
activities in string theory.  Chiral fermions on higher-dimensional
membranes are in much the same spirit as the domain-wall fermion
approach.

So, well into its third decade, lattice gauge theory remains a
thriving industry.  While dominated by numerical work, the field is
considerably broader.  The unsolved problems, particularly with
fermionic fields, show that despite the maturity of the subject we
still need new ideas!

\bigskip
{\noindent \bf Acknowledgement:} This manuscript has been authored
under contract number DE-AC02-98CH10886 with the U.S.~Department of
Energy.  Accordingly, the U.S. Government retains a non-exclusive,
royalty-free license to publish or reproduce the published form of
this contribution, or allow others to do so, for U.S.~Government
purposes.

\begin{figure}
\epsfxsize .25\hsize
\centerline{\epsffile {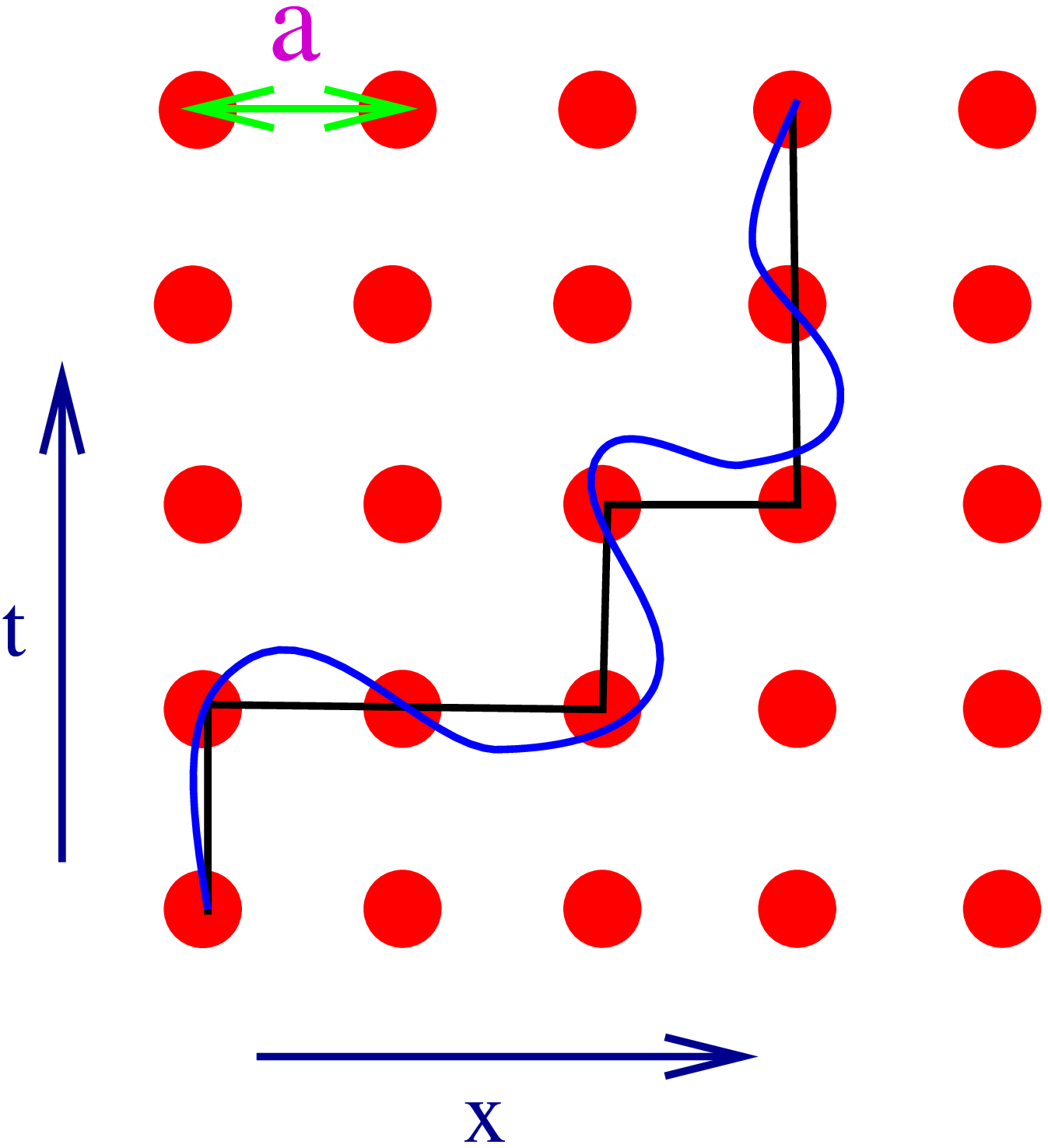}}
\medskip
\caption {Lattice gauge theory approximates quark world-lines by
sequences of hops through a four-dimensional lattice.  The spacing
must be extrapolated to zero for physical results.}
\end{figure}

\begin{figure}
\epsfxsize .25\hsize
\centerline{\epsffile {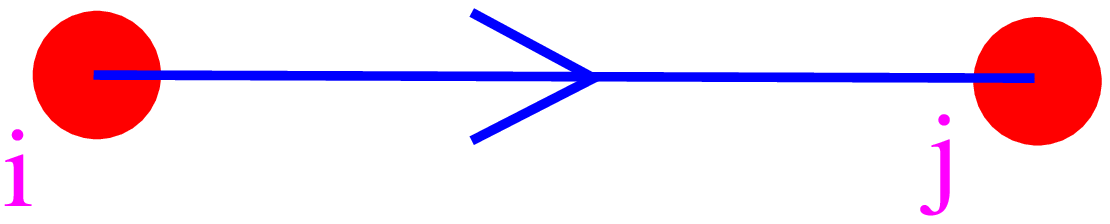}}
\medskip
\caption {On each link connecting two nearest neighbors is a
3 by 3 unitary matrix $U_{ij} \in SU(3)$. }
\end{figure}

\begin{figure}
\epsfxsize .2\hsize
\centerline{\epsffile {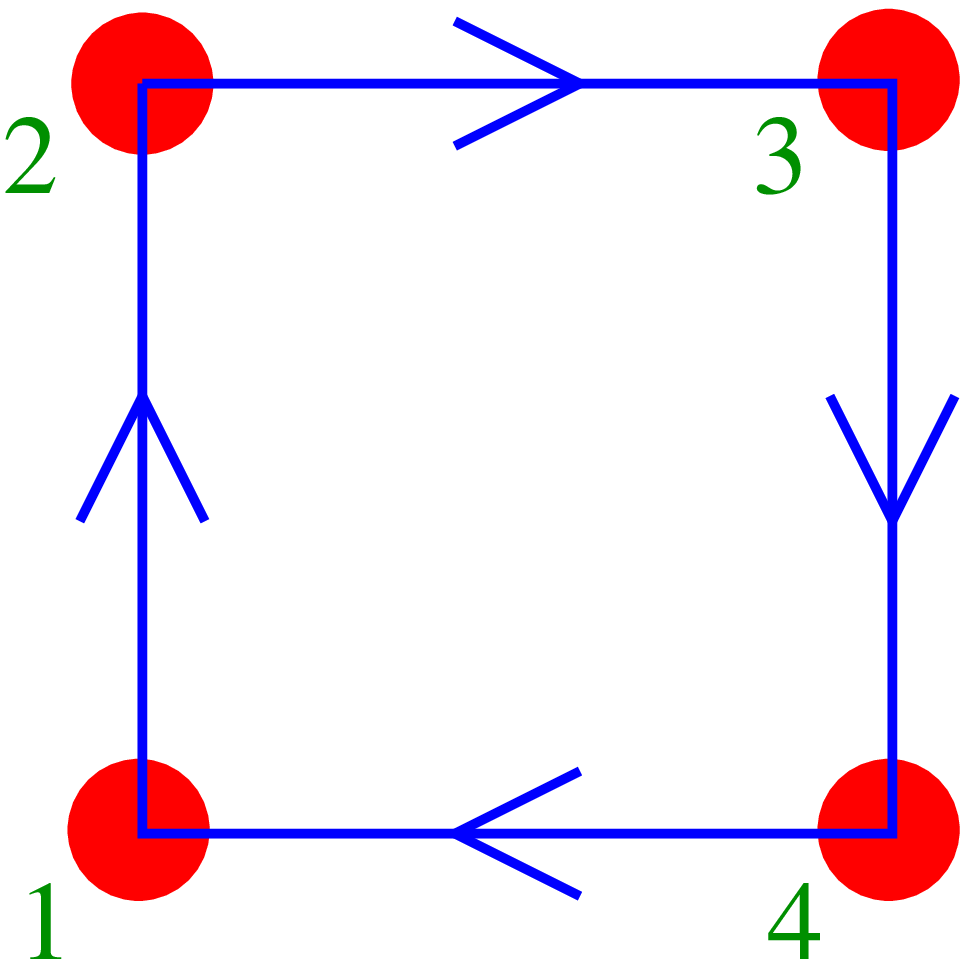}}
\medskip
\caption {Multiplying the phase factors around a plaquette gives the
flux through that plaquette.  This is used to construct the action.
}
\end{figure}

\begin{figure}
\epsfxsize .5\hsize
\centerline {\epsfbox{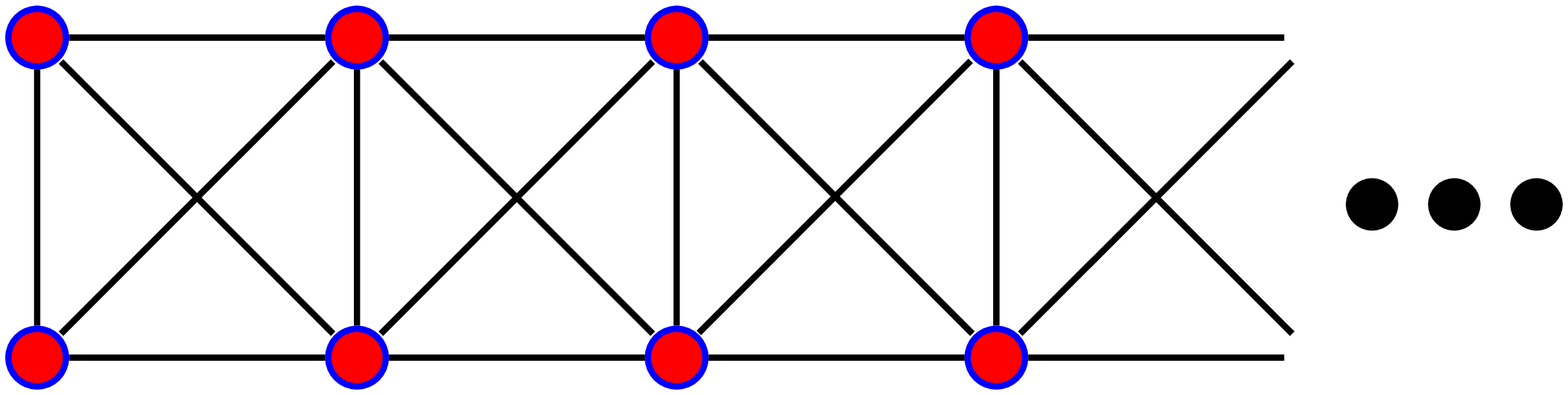}}
\medskip
\caption {The basic ladder molecule discussed in the text.}
\end{figure}

\begin{figure}
\epsfxsize .5\hsize
\centerline {\epsfbox{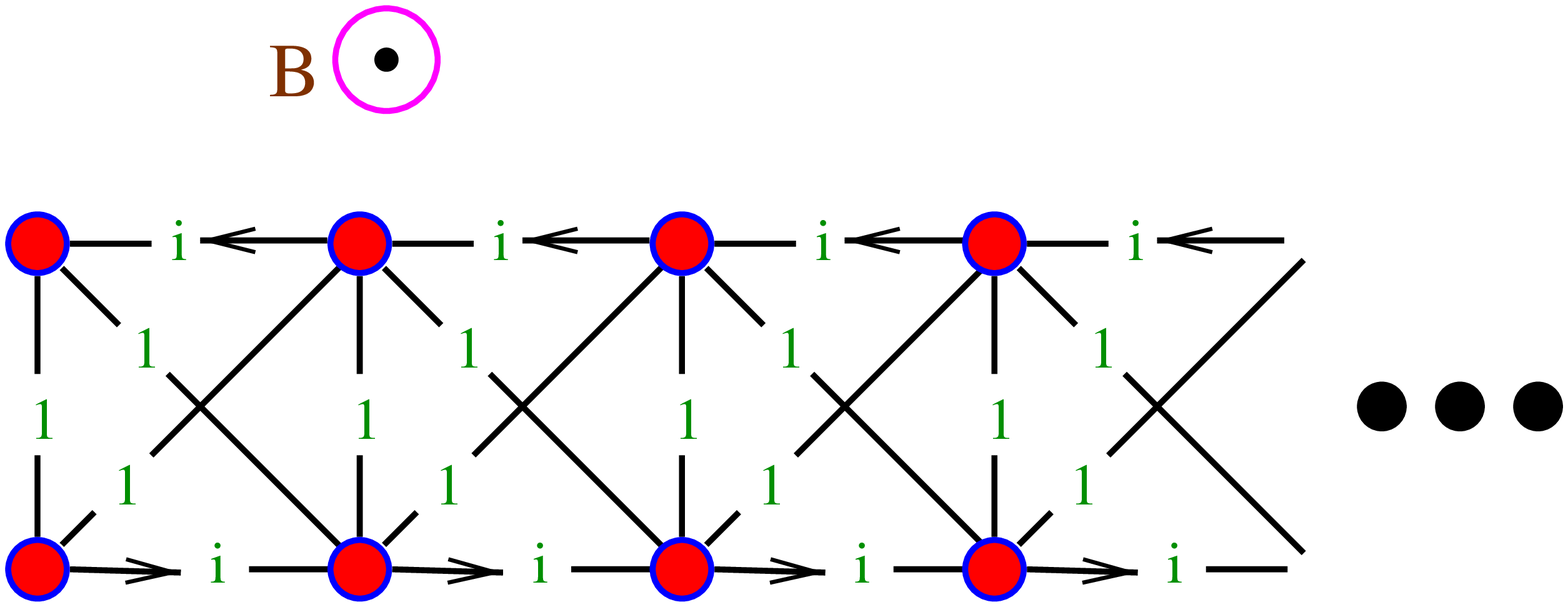}}
\medskip
\caption {Applying a field of one half flux unit per plaquette gives
phases on the bonds.  One convenient gauge choice is shown here.}
\end{figure}

\begin{figure}
\epsfxsize .5\hsize
\centerline {\epsfbox{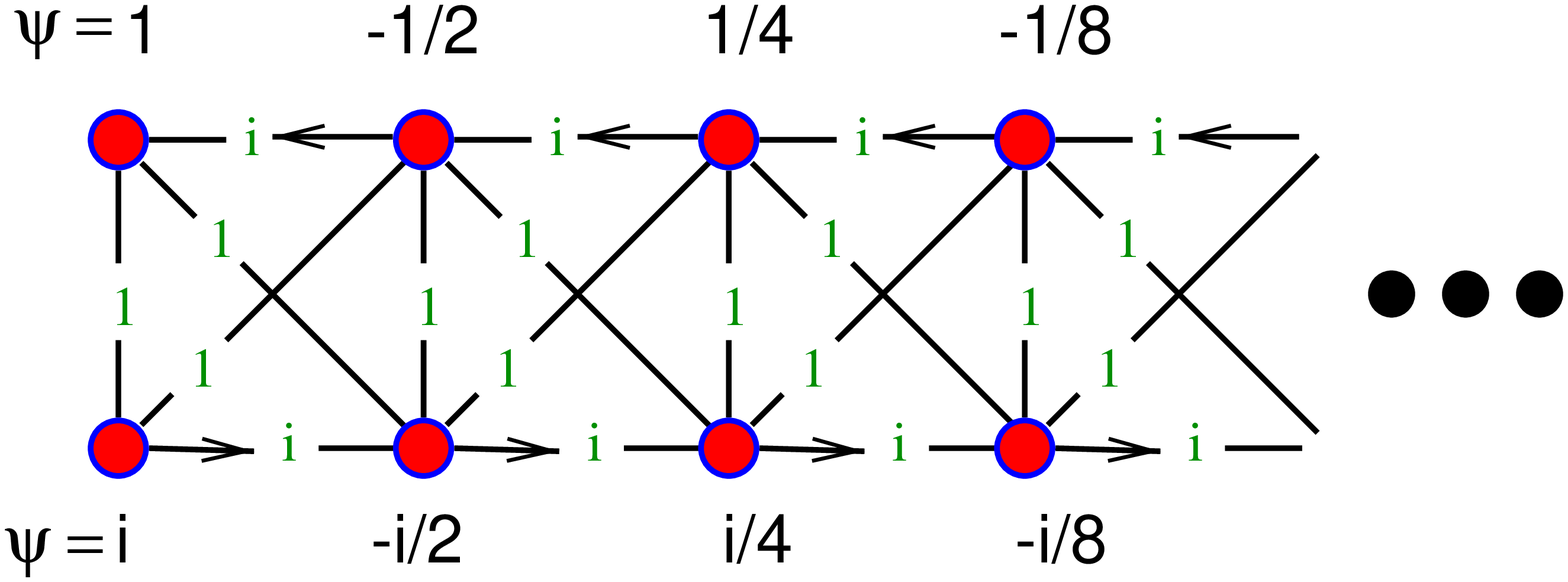}}
\medskip
\caption {The wave function for a zero energy state bound on the end of
the ladder molecule.}
\end{figure}

\begin{figure}
\epsfxsize .4\hsize
\centerline {\epsfbox{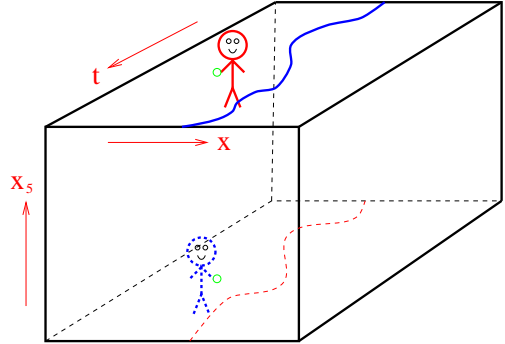}}
\medskip
\caption {For domain-wall fermions, our four-dimensional
world is interpreted as a surface on a five-dimensional manifold.
Zero modes on this surface are the physical quark degrees of freedom.
}
\end{figure}

\begin{figure}
\epsfxsize .5\hsize
\centerline {\epsfbox{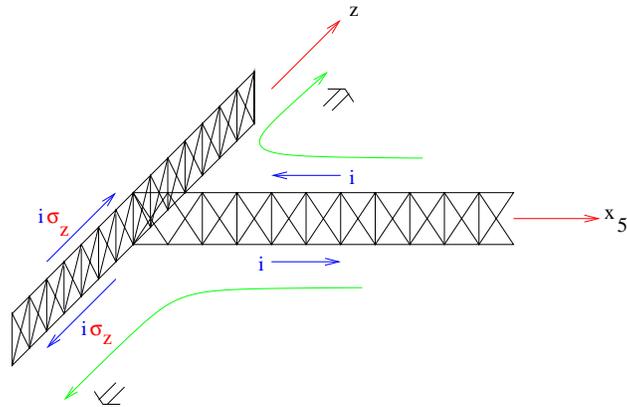}}
\medskip
\caption {A device constructed by joining ladder molecules.  Inserting
Pauli matrices in the spatial directions creates a helicity filter.  }
\end{figure}

\bigskip
\begin{figure}
\epsfxsize .25\hsize
\centerline {\epsfbox{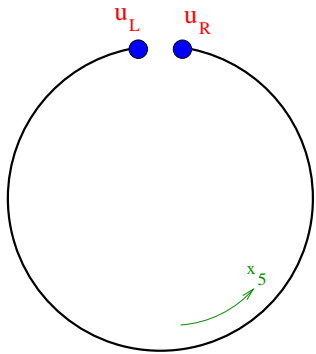}}
\medskip
\caption {A compact fifth dimension must be cut in order to generate
the chiral zero modes of the domain-wall formalism.}
\end{figure}

\begin{figure}
\epsfxsize .25\hsize
\centerline {\epsfbox{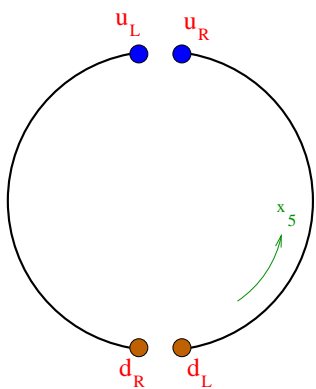}}
\medskip
\caption {Cutting the compact fifth dimension twice
gives two flavors of fermion.  With the identifications here,
one flavored chiral symmetry is exact, even when the lattice
spacing and the size of the fifth dimension are finite.}
\end{figure}

\begin{figure}
\epsfxsize .4\hsize
\centerline {\epsfbox{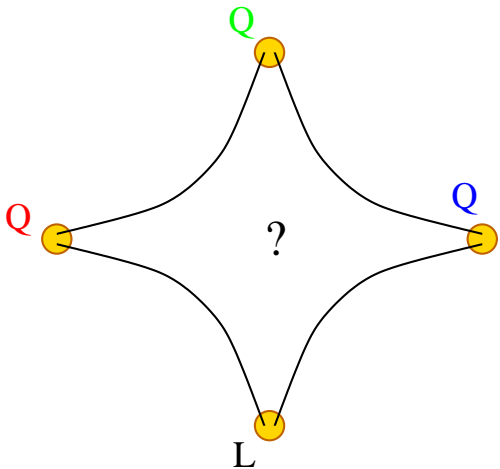}}
\medskip
\caption {Perhaps all fermions in a generation are special modes of a
single higher-dimensional field.  Here the three quark fields might
represent different values of the internal $SU(3)$ symmetry, and $L$
could represent a lepton from the same family.}
\end{figure}

\begin{figure}
\epsfxsize .5\hsize
\centerline {\epsfbox{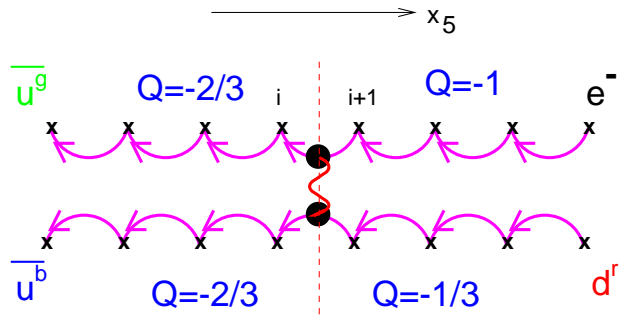}}
\medskip
\caption {One specific model with three quark colors and a lepton all
being manifestations of a single five-dimensional fermion field.  }
\end{figure}

\end{document}